\documentclass{elsart}
\usepackage{epsfig}
\usepackage{cite}

\newcommand{\et}{\mbox{$\eta$}}
\newcommand{\po}{\mbox{$\pi^\circ$}}

\newcommand{\pim}{\mbox{$\pi^-$}}
\newcommand{\pip}{\mbox{$\pi^+$}}

\newcommand{\pppipipi}{\mbox{$pp\to pp\pi^{+}\pi^{-}\po$}}
\newcommand{\pnpipipi}{\mbox{$pp\to pn\pi^{+}\pi^{+}\pi^{-}$}}
\newcommand{\pppopopo}{\mbox{$pp\to pp\po\po\po$}}
\newcommand{\pptpi}{\mbox{$pp\to pp \pi\pi\pi$}}
\newcommand{\ppe}{\mbox{$pp\rightarrow pp\et$}}

\newcommand{\E}[1]{\mbox{$\times$10$^{#1}$}}

\begin{document}
\begin{frontmatter}
\title{The $pp\to pp \pi\pi\pi$ reaction channels in the threshold region.}
\collab{CELSIUS-WASA Collaboration}
\runauthor{C. Pauly et al.}
\author[Hamburg]{C.~Pauly},
\author[UU]{M.~Jacewicz},
\author[UU]{I.~Koch},
\author[Tuebingen]{M.~Bashkanov},
\author[Dubna]{D.~Bogoslawsky},
\author[TSL]{H.~Cal\'en},
\author[UU]{F.~Capellaro},
\author[Tuebingen]{H.~Clement},
\author[Hamburg]{L.~Demir\"ors}, 
\author[TSL]{C.~Ekstr\"om},
\author[TSL]{K.~Fransson},
\author[UU]{L.~Gustafsson},
\author[UU]{B.~H\"oistad},
\author[Dubna]{G.~Ivanov},
\author[Dubna]{E.~Jiganov},
\author[UU]{T.~Johansson}, 
\author[UU]{S.~Keleta},
\author[UU]{S.~Kullander},
\author[TSL]{A.~Kup\'s\'c\thanksref{cor}},
\thanks[cor]{Corresponding author. {\sl E-mail address:} kupsc@tsl.uu.se}
\author[Dubna]{A.~Kuznetsov},
\author[TSL]{P.~Marciniewski},
\author[Tuebingen]{R.~Meier},
\author[Dubna]{B.~Morosov},
\author[Juelich]{W.~Oelert},
\author[Dubna]{Y.~Petukhov},
\author[UU]{H.~Pettersson},
\author[Dubna]{A.~Povtorejko},
\author[TSL]{R.J.M.Y.~Ruber},
\author[UU]{K.~Sch\"onning},
\author[Hamburg]{W.~Scobel},
\author[Tuebingen]{T.~Skorodko},
\author[Novosibirsk]{B.~Shwartz},
\author[ITEP]{V.~Sopov},
\author[SINS]{J.~Stepaniak},
\author[ITEP]{V.~Tchernyshev},
\author[UU]{P.~Th\"orngren Engblom},
\author[Dubna]{V.~Tikhomirov},
\author[WU]{A.~Turowiecki},
\author[Tuebingen]{G.J.~Wagner},
\author[UU]{U.~Wiedner},
\author[UU]{M.~Wolke},
\author[Tsukuba]{A.~Yamamoto},
\author[SINS]{J.~Zabierowski},
\author[UU]{J.~Z{\l}oma\'nczuk}
\address[Hamburg]{Institut f\"ur Experimentalphysik der Universit\"at Hamburg,
Hamburg, Germany}
\address[UU]{Institutionen f\"or k\"arn- och partikelfysik, Uppsala University,
 Uppsala, Sweden}
\address[Tuebingen]{Physikalisches Institut der Universit\"at T\"ubingen, 
T\"ubingen, Germany}
\address[Dubna]{Joint Institute for Nuclear Research, Dubna, Russia}
\address[TSL]{The Svedberg Laboratory, Uppsala, Sweden}
\address[Juelich]{Institut f\"ur Kernphysik, Forschungszentrum J\"ulich, 
J\"ulich, Germany}
\address[Novosibirsk]{Budker Institute of Nuclear Physics, Novosibirsk, Russia}
\address[ITEP]{Institute of Theoretical and Experimental Physics, 
Moscow, Russia}
\address[SINS]{Soltan Institute of Nuclear Studies, Warsaw and Lodz, Poland}
\address[WU]{Institute of Experimental Physics, Warsaw, Poland}
\address[Tsukuba]{High Energy Accelerator Research Organization, 
Tsukuba, Japan}

\begin{abstract}
The cross section for prompt neutral and charged three pion production
in $pp$ interactions was measured  at excess energies in the range 160
-- 217  MeV.   That comprises  the  first  measurement  of the  $pp\to
pp\pi^\circ\pi^\circ\pi^\circ$ reaction and the direct comparison with
the  $pp\to   pp\pi^+\pi^-\pi^\circ$  process.   The   experiment  was
performed above  the $\eta$ meson  production threshold and  the cross
section  could be  directly normalized  to  the cross  section of  the
$pp\to pp\eta$ reaction, with the $\eta$ decaying into 3 pions.  Since
the  same  final  states  are  selected, the  measurement  has  a  low
systematical  error.  The measured  cross section  ratio $\sigma(pp\to
pp\pi^+\pi^-\pi^\circ)/  \sigma(pp\to  pp\pi^\circ\pi^\circ\pi^\circ)$
is compared  to predictions of  dominance of different isobars  in the
intermediate state.

\end{abstract}

\end{frontmatter}

\section{Introduction}

Production  of three  pions in  proton-proton interactions,  where the
pions do  not originate from  decays of narrow meson  resonances (like
$\eta$ or  $\omega$), has not  received proper attention  yet, neither
experimentally nor  theoretically. Experimentally there  are only very
few data on the \pppipipi\ and \pnpipipi\ reactions and no data on the
other reaction channels: \pppopopo, $pp\to pn\pi^{+}\po\po$ and $pp\to
nn\pi^{+}\pi^{+}\po$.   On the  theory  side there  exists a  complete
microscopic  model   covering  all  reaction  channels   of  two  pion
production in p$N$ interactions \cite{Alvarez-Ruso:1997mx}, but so far
no such  models have been developed  for the three pion  case.  In the
isobar model  the process \pptpi\  should proceed by an  excitation of
one  or  two  baryon  resonances  followed by  the  subsequent  decays
\cite{Sternheimer:1961} and in the  low energy region a mechanism with
a   simultaneous  excitation  of   $N^*$  and   $\Delta(1232)  P_{33}$
resonances is expected  to dominate.  The $N^*$ involved  has to decay
into $N\pi\pi$ and therefore the lowest lying Roper ($N(1440) P_{11}$)
and $N(1520) D_{13}$ resonances could be considered.  There is however
a  significant  difference between  their  decay  pattern.  The  Roper
decays predominantly  into $\Delta\pi$  whereas for the  $N(1520)$ the
$N\rho$  channel is equally  important \cite{Vrana:1999nt,Yao:2006px}.

The  influence of  the resonances  in  the intermediate  state can  be
studied in the  invariant mass distributions of the  subsystems of the
outgoing protons and pions.  Such studies were done for the \pppipipi\
and \pnpipipi\  reactions in  bubble chamber experiments  performed at
higher energies (beam kinetic energies  of 4.15 GeV and 9.11 GeV) with
up  to   thousand  events  \cite{Alexander:1967aa,  Colleraine:1967aa,
Almeida:1969bv}.  In these studies, only  one or two of the pions were
considered to  originate from decays of $N^*$  or $\Delta$ resonances.
However,  such  kind  of  analysis  is  complicated  and  may  not  be
conclusive due to  many possible scenarios and due  to large widths of
the involved  resonances.  A simple  starting point for  analyzing the
three pion production close to threshold is to assume a constant value
for the  matrix element.  The dependence  of the cross  section on the
beam energy is then given by  the phase space volume divided by a flux
factor.  For  a proper description  of the reactions in  the threshold
region it is required to take into account the final state interaction
(FSI)    between    the    outgoing    protons    \cite{Faeldt:1996na,
Hanhart:2003pg}.  The $N\pi$  FSI on the other hand  is expected to be
negligibly small \cite{Gell-Mann:1954wr}.  A hint about the production
mechanism of  the three pions could  then be obtained  by studying the
ratio of the cross sections for the different charge states. The three
pion production amplitude $NN\to  NN\pi\pi\pi$ can be written in terms
of  the isospin amplitudes  $M_{T_iT_{3\pi}T_f}$, where  $T_i$ ($T_f$)
denotes the initial (final) isospin of the nucleon pair and $T_{3\pi}$
denotes  the  isospin  of  the  produced  pion  triplet.   It  is  now
straightforward to show that
\[
\begin{array}{lll}
\sigma(\pppipipi)&\propto&\frac{1}{30}|M_{121}|^2+  \frac{1}{5}|M_{111}|^2+
\frac{1}{6}|M_{101}|^2\\  
\sigma(\pppopopo)&\propto&\frac{1}{20}|M_{111}|^2\\
\sigma(\pnpipipi)&\propto&\frac{1}{20}|M_{121}|^2+\frac{3}{20}|M_{111}|^2+
\frac{3}{10}|M_{110}|^2,
\end{array}
\]
where  cross  terms between  $M_{121}$  and  $M_{101}$ amplitudes  are
neglected.   In the  simple statistical  approach \cite{Fermi:1950jd},
with  all  amplitudes  $M_{T_iT_{3\pi}T_f}$  put  equal,  one  obtains
$\sigma(\pppipipi):    \sigma(\pppopopo):   \sigma(\pnpipipi)=8:1:10$.
There     were      so     far     experimental      data     on     $
\sigma(\pppipipi):\sigma(\pnpipipi)$  only. At 2.0  GeV this  ratio is
1:2.53$\pm$0.46   \cite{Pickup:1962aa}   and  at   2.85   GeV  it   is
1:1.59$\pm$0.27 \cite{Hart:1962aa}. This suggests a deviation from the
statistical   approach,  especially   at   lower  energies.    Similar
conclusions  can  also  be  drawn  from  experiments  on  double  pion
production  at excess  energies below  100  MeV \cite{Johanson:2002hs}
where the  statistical approach fails.   We discuss the  cross section
ratios  under  different  assumptions  about the  dominating  reaction
mechanism in section 3 of this paper.

With  a 4$\pi$ facility  such as  WASA \cite{Adam:2004ch},  aiming for
measurements of decays  of $\eta$ and $\eta'$ mesons  produced in $pp$
interactions, the understanding of  the \pptpi\ reactions becomes very
important as  they constitute a  severe background for the  studies of
$\eta$ and $\eta'$ decays into  three pions.  Those decays provide key
ingredients  for determining  the  ratios of  the  light quark  masses
$m_u/m_s$  and  $m_d/m_s$ \cite{Gross:1979ur,Leutwyler:1996qg},  since
the  decay widths  are  proportional to  the  $u$ and  $d$ quark  mass
difference squared.  In order to estimate the effect of the background
for the decay experiments, the properties of the \pptpi\ reactions for
beam  proton energies  from  1.254 GeV  (corresponding  to the  $\eta$
production  threshold)   to  3  GeV  are  important.    There  are  no
experimental points  for the cross section of  the \pppopopo\ reaction
in that energy range, and  only three points for \pppipipi\ all coming
from          old          bubble         chamber          experiments
\cite{Pickup:1962aa,Hart:1962aa,Eisner:1965}.       The     experiment
performed at  the lowest energy,  1.48~GeV, identified a  single event
\cite{Eisner:1965}  (corresponding  to $20\mu  b$).   Moreover in  the
bubble chamber  experiments, the prompt three pion  production was not
separated from the $\eta \rightarrow \pi\pi\pi$ decays.

\section{Measurement}

The  analysis is based  on data  collected with  the WASA  facility at
CELSIUS  \cite{Zabierowski:2002ah}. The  target system  provides small
($\phi  \approx 30\  \mu$m) hydrogen  pellets that  interact  with the
circulating proton beam. The  protons have nominal kinetic energies of
1.30, 1.36 and 1.45~GeV (corresponding to excess energies, $Q$, in the
center of mass  system around 200 MeV for  the \pptpi\ reactions). The
integrated   luminosities   at   each   energy   were:   27~nb$^{-1}$,
414~nb$^{-1}$  (data on  \pppipipi\  reaction are  based  on a  sample
corresponding  to 80~nb$^{-1}$)  and 221~nb$^{-1}$  respectively.  The
WASA detector  system consists of a multilayer  forward detector (FD),
for measurement of the outgoing  protons scattered in an angular range
of  2.5--18$^{\circ}$,  and  a  central detector  (CD)  containing  an
electromagnetic calorimeter and a drift chamber/solenoid for measuring
the produced mesons  and their decay particles in  an angular range of
20--140$^{\circ}$.  The  experimental method for  extracting the cross
section  of  the  \pptpi\  reaction,  relies  on  normalizing  to  the
simultaneously  measured   $pp  \rightarrow  pp\eta$   reaction,  with
subsequent decay of the eta into $\pip\pim\po$ (branching ratio 22.7\%
\cite{Yao:2006px}) or  into $\po\po\po$ ($32.5\%$).  In  this way, the
reference reaction has the same particles in the final state, and most
of the efficiency corrections cancel.   The final state is selected by
the requirement that two protons are detected in the forward detector.
In addition the selected events  should have six $\gamma$ hit clusters
(with energy depositions  of at least 20 MeV)  in the calorimeter, for
the $3 \pi^{\circ}$ case, or  two $\gamma$ hit clusters and two tracks
with opposite bending  in the central detector drift  chamber, for the
$\pi^+\pi^-\pi^{\circ}$ case.  After  identification of the two proton
tracks in  the FD no additional  cuts on the  proton-proton system are
applied.   The specific  cuts  for neutral  and  charged CD  particles
systems aim  to select  a clean  3$\pi$ final state.   In case  of the
$pp\po\po\po$ channel, already the  requirement of the six neutral hit
clusters  in  the CD  results  in a  fairly  clean  data sample.   The
$\pi^+\pi^-\pi^0$ sample was obtained  by requiring the invariant mass
of the two  photons to be located in the $\pi^0$  mass region, and the
missing mass  of the two  protons plus the  two photons to  be greater
than two pion masses.

Two pion production, $pp \rightarrow pp \pi\pi$, is the major physical
background which has  to be considered in the analysis  in view of its
much higher cross section.  The  final state can be misidentified as a
three pion  event only in case  of two additional hit  clusters in the
calorimeter, due  to cluster split-offs, pile-up or  noise.  To reject
the background  strict time  cuts are applied,  as well  as additional
kinematical conditions -- such  as reconstruction of the total energy.
The final contribution of the background  is on the per cent level, in
agreement  with   expectations  from  the   Monte  Carlo  simulations.
Therefore  the uncertainty  in  the  exact value  of  the total  cross
section  for  the  two pion  channels,  as  well  as in  the  reaction
mechanism, only gives a minor contribution to the systematical error.

 \begin{figure}
   \begin{minipage}{0.5\textwidth}
    \includegraphics[width=\textwidth]{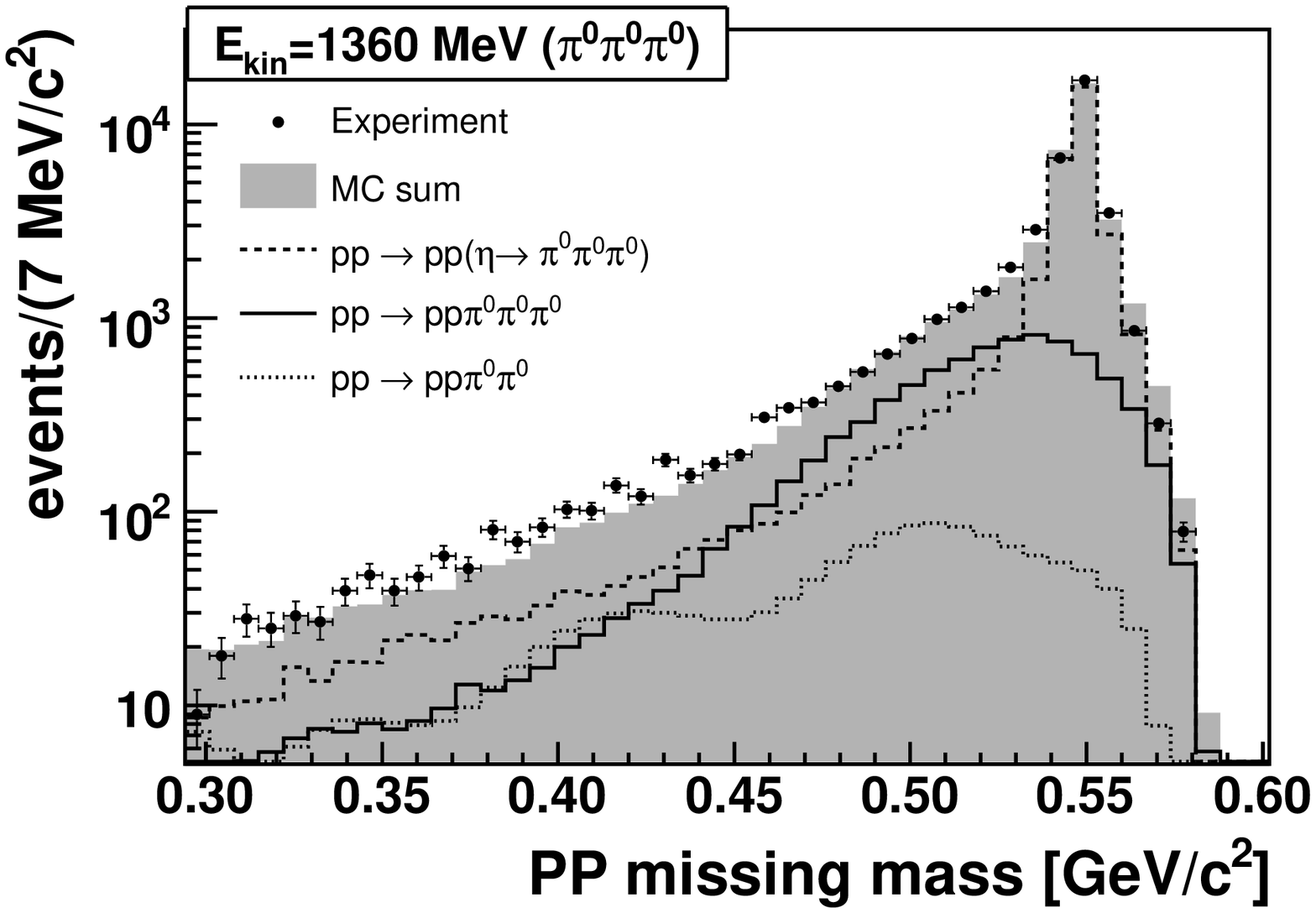}
  \end{minipage}\hfill\begin{minipage}{0.5\textwidth}
    \includegraphics[width=\textwidth]{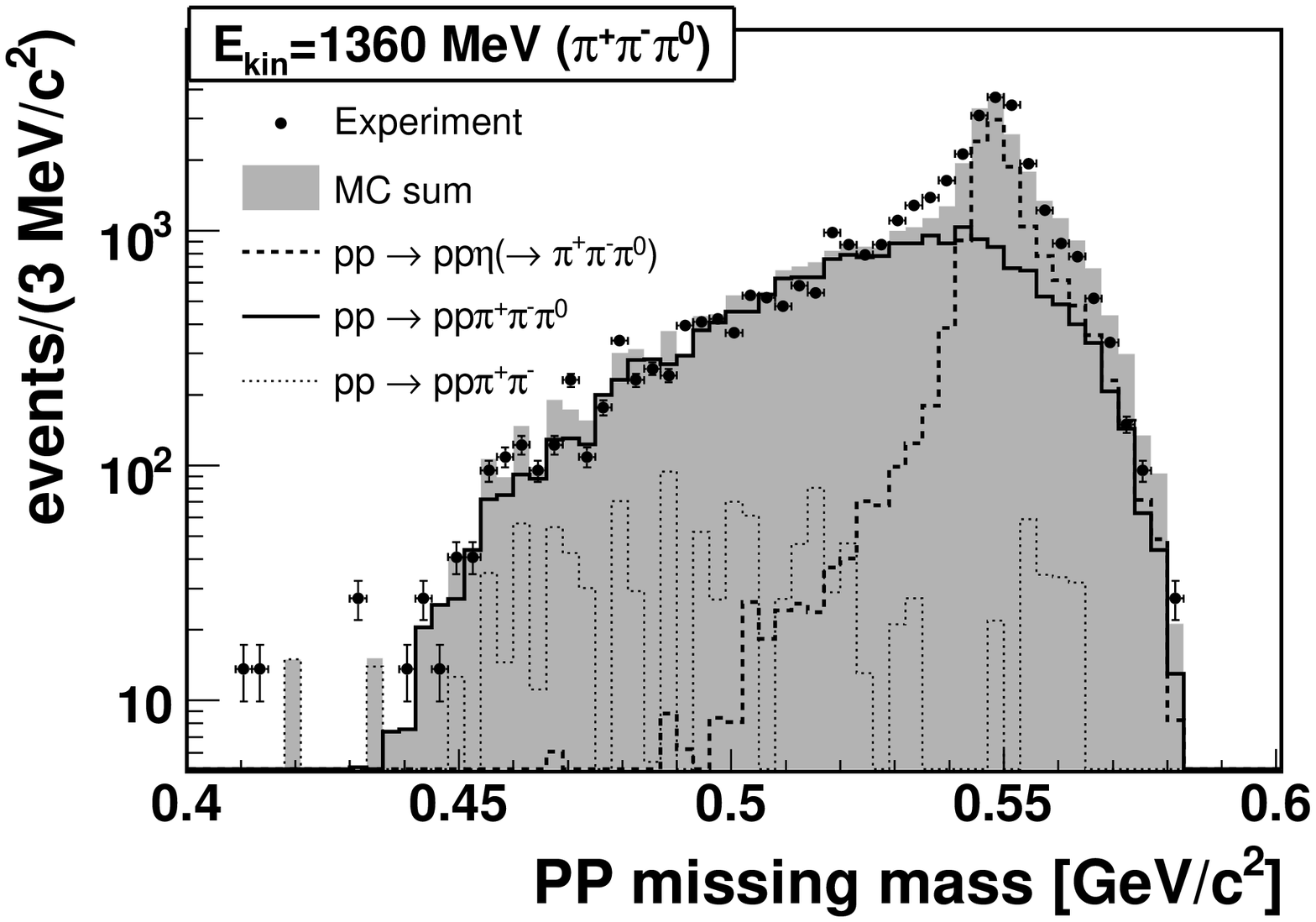}
  \end{minipage}
   \begin{minipage}{0.5\textwidth}
    \includegraphics[width=\textwidth]{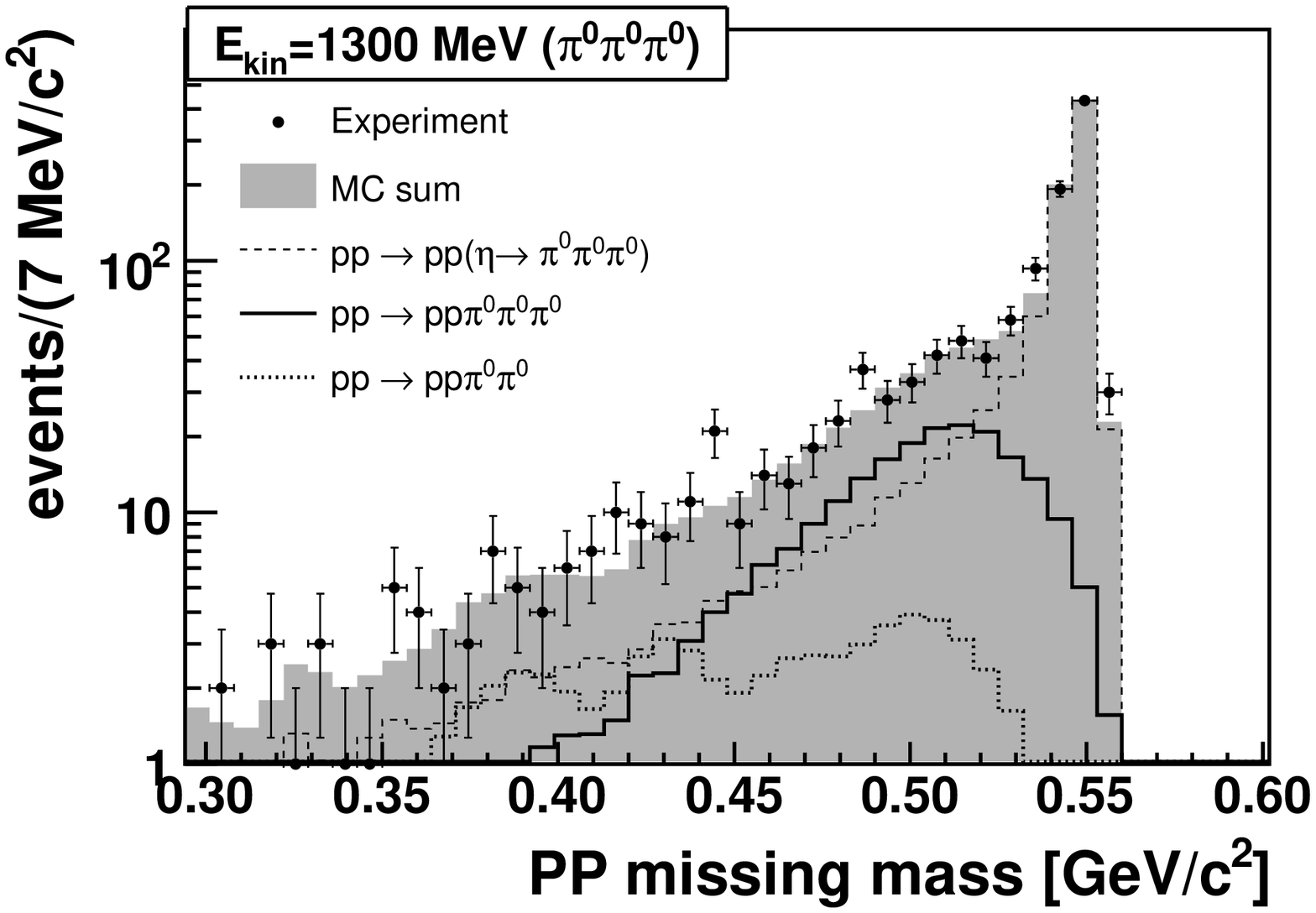}
  \end{minipage}\hfill\begin{minipage}{0.5\textwidth}
    \includegraphics[width=\textwidth]{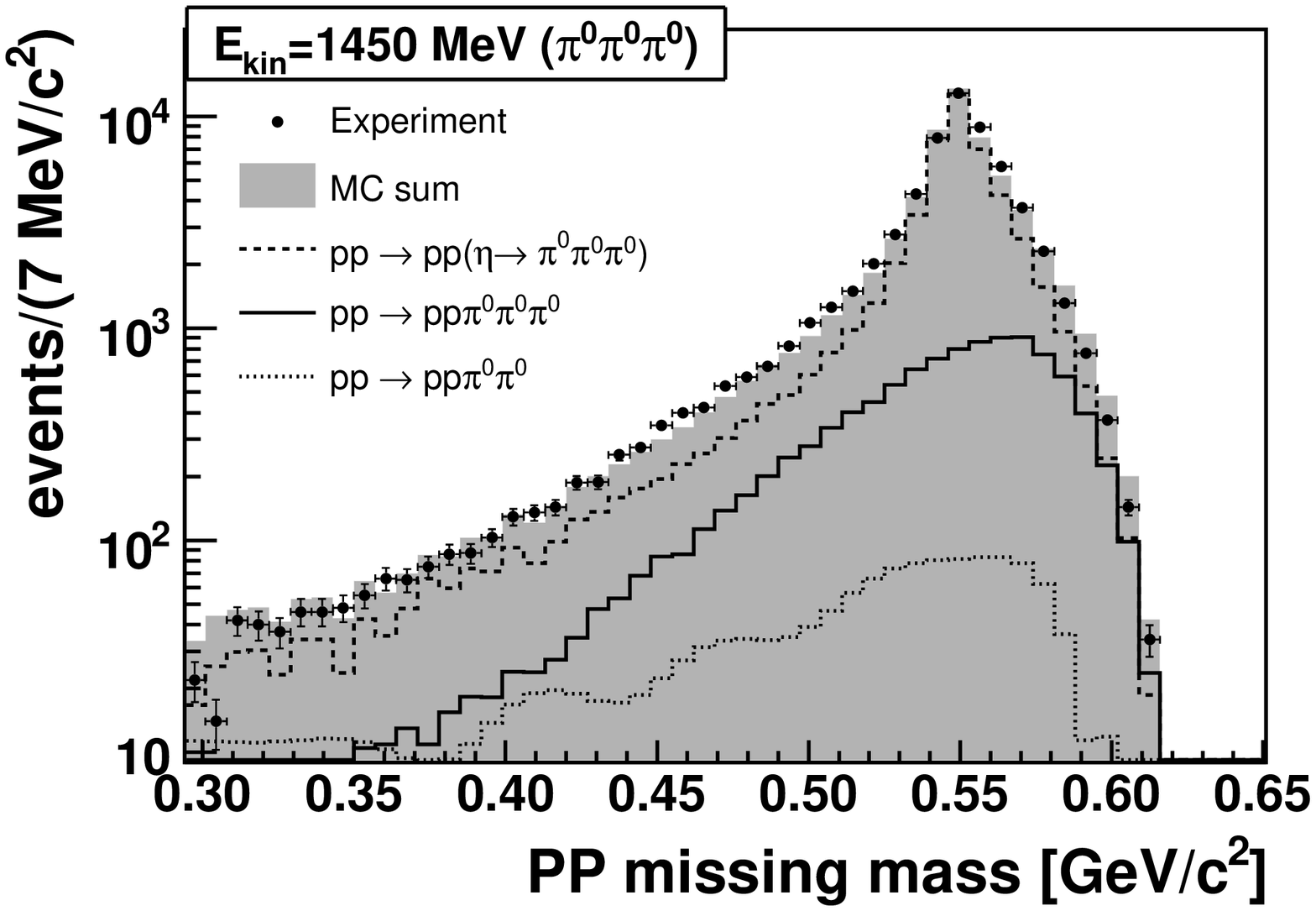}
  \end{minipage}

    \caption[pi0]{\label{fig:csection} Experimental $pp$ missing mass
    distributions for $pp\pi\pi\pi$ final states with fit of the Monte
    Carlo distributions for 2$\pi$, 3$\pi$ and $\eta$ production.  }
\end{figure}

The derivation of the cross sections for the $pp\rightarrow pp3\pi$ 
reaction is based  on the  fact that the  resolution in  the $pp$
missing mass determination is very good (4 MeV/c$^2$ FWHM at 1.30 GeV) 
and that the missing mass distribution is practically insensitive to the 
mechanism of the $pp\to pp\eta$ reaction. 

After  the  selection of  the  $pp\pi\pi\pi$  final  states, the  $pp$
missing mass  distribution is  constructed from experimental data, and
from Monte Carlo simulated data of the reaction channels \pptpi, $pp\to
pp(\eta\to  3\pi)$,  and $pp\to  pp  \pi\pi$  as  the main  background
contribution.   The  individual Monte  Carlo  distributions were  then
fitted to  the measured spectrum,  with the \pptpi\ to  $pp\to pp\eta$
ratio as parameter. The $pp\to pp \pi\pi$  contribution was fixed
relative  to the  $pp\to pp\eta$  channel based  on the  cross section
values listed in table \ref{tab:0}.  Fig.~\ref{fig:csection} shows the
result of the  best fit at the three beam  energies for the \pppopopo\
reaction and at 1.36 GeV for the \pppipipi\ reaction.  Several studies
were done in order to determine uncertainties and check consistency:
\begin{itemize}
\item For  1.36 GeV,  data from two well separated  run periods were  
analyzed and gave consistent results.
\item The kinematical cuts were varied.
\item  Application  of  a  kinematical  fit to  the \pptpi\  reaction  and
verification of  the $\chi^2$ distribution (7 constrains  (7C) fit for
the \po\po\po\ and 5C fit for the \pip\pim\po).
\item  Variation of the  nominal  beam  energy  and of  the  energy
resolution of the detectors assumed in the analysis.
\end{itemize}

\begin{figure}

  a)\includegraphics{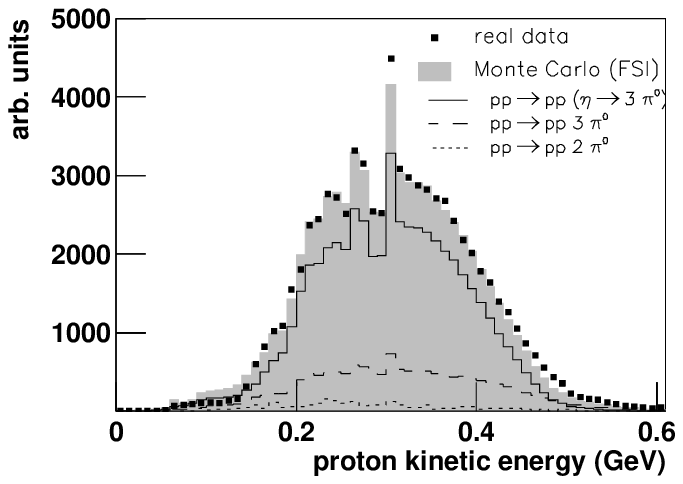}
  b)\includegraphics{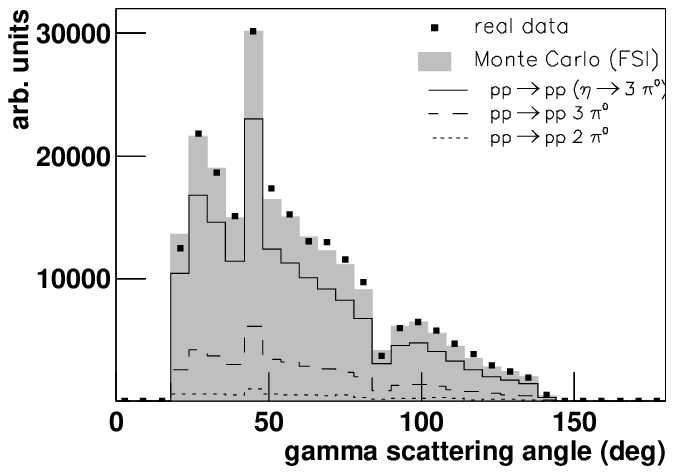}                       \\
  c)\includegraphics{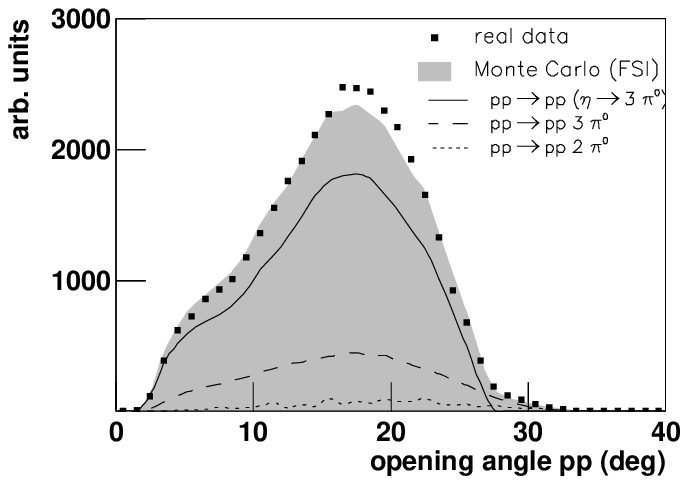}

\caption[MC]{\label{fig:MCquality}   Comparison of experimental distributions 
in the lab system of a) proton kinetic energies, b) $\gamma$ reaction angles, 
and c) $pp$ opening angle for beam proton kinetic energy  1.36 GeV 
projectiles with Monte Carlo simulation of the  
$pp3\pi^{\circ}, pp2\po$ and 
$pp(\eta\to 3\po)$ best fit including $pp$ FSI.}
\end{figure}

The applied method to obtain the cross section ratio relies heavily on
the  Monte  Carlo  simulations.   Their accuracy  is  demonstrated  in
fig.~\ref{fig:MCquality} by comparing  distributions of some variables
with  the result  of the  simulation,  taking into  account all  three
mentioned reaction channels.  The  normalization factors for the three
channels are  the same for  each distribution, and were  obtained from
the fit of the pp missing mass distribution (fig.~\ref{fig:csection}).
The simulation  describes well detector effects such  as the structure
at 0.3  GeV, which  is due to  an ambiguity in  distinguishing stopped
protons  from   punch  through  protons   in  the  FD.    The  angular
distribution  of the  $\gamma$s  shows two  distinct, well  reproduced
structures.   Near 40$^{\circ}$  the  binning changes  to account  for
different sizes  of the CsI  scintillators in the CD;  at 90$^{\circ}$
the detection  efficiency is reduced by the  pellet target components.
The major  geometric acceptance limitation  is imposed by  the angular
acceptance   for   protons   (3$^{\circ}$--17$^{\circ}$).   The   good
agreement between simulated and real data both in the gamma and proton
scattering angle distribution indicates that the acceptance correction
is under good control.

\begin{figure}
  \begin{minipage}[t]{0.5\textwidth}
\end{minipage}
   \begin{minipage}[t]{0.5\textwidth}
    \includegraphics[width=\textwidth]{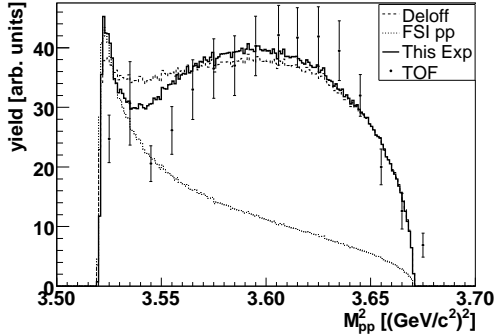}
   \end{minipage}

\caption[ang]{\label{fig:angle}  
Proton-proton  invariant mass squared at 1.36  GeV.  The solid
line shows  the parametrization from this experiment.  The dashed line
shows  the  model prediction  from  A.  Deloff  \cite{Deloff:2003te} tuned  to
describe the TOF data \cite{Abdel-Bary:2002sx} and dotted line the $pp$ FSI.}
\end{figure}

In order to extract the total cross section, a model has to be used to
extrapolate  data  outside  of   the  measured  range  of  the  proton
scattering  angles.   For the  $pp  \rightarrow  pp\eta$ reaction  the
largest deviation from the phase space behavior is expected due to the
strong $pp$ final state interaction.  We use a full calculation of the
$pp$   FSI   of   A.~Deloff   \cite{Deloff:2003te}   with   the   Reid
$NN$-potential,  which yields  a perfect  reproduction of  the $^1S_0$
phase  shifts  for  relative  $pp$  momenta  up  to  0.3~GeV/c.   This
calculation reproduces the $pp$  angular distribution much better, but
not yet  the squared missing mass  $(M^2_{pp})$ distributions obtained
for   $Q   =$  15.5~MeV   \cite{Moskal:2003gt}   and   $Q  =$   41~MeV
\cite{Abdel-Bary:2002sx}.     Agreement   was    then    obtained   in
\cite{Deloff:2003te}  by  expanding the  production  amplitude in  the
$\eta$  momentum  relative  to  the $pp$  pair.   The  proportionality
constant was optimized for a  best fit of the $M^2_{pp}$ distribution,
cf.   fig.~\ref{fig:angle}.   Application  of  this  approach  to  the
$M^2_{pp}$ distribution of this work  yields a very similar result for
$Q  =$  41~MeV  (see  fig.~\ref{fig:angle})  and  $Q  =$  17~MeV  (not
shown). This fit at the same time gives a good description of the $pp$
opening angle distribution (fig.~\ref{fig:MCquality}c), where the $pp$
FSI  leads  to an  enhancement  at the  lowest  angles.   For the  $pp
\rightarrow pp3\pi$,  the outgoing protons can  have scattering angles
up to 30$^\circ$  at those beam energies.  This  means that extraction
of the total cross section involves some extrapolation into unmeasured
kinematical  regions.  However, it  was checked  that {\em  e.g.}  the
inclusion of  a broad resonance in  the $N\pi$ system  in the reaction
mechanism changes the total acceptance by a few per cent only.

\begin{table}
\caption[tab1]{\label{tab:0}  Total  cross sections  for two
pion  and  eta   production  used  in  the  analysis.   The  data  for
$\pi^o\pi^o$ production are  extracted from bubble chamber experiments
\cite{Shimizu:1982dx,Eisner:1965}  and from CELSIUS/WASA  \cite{Koch:2005aa}.
 The  data for
$\pi^+\pi^-$ production  are from ref. \cite{Brunt:1970tg}. The  data for eta
production     are      from     CELSIUS,     COSY      and     Saclay
\cite{Calen:1996mn,Chiavassa:1994ru,Moskal:2003gt,Hibou:1998de}.
   Some   values   were   obtained   by
interpolation to the energies used in this experiment.}
\begin{tabular}{rrrr}
\hline Beam kinetic& pp$\eta$     & pp$\pi^+\pi^-$&pp$\pi^0\pi^0$ \\ energy
$[$GeV$]$          &[$\mu$b]      &$[\mu$b$]$&[$\mu$b]\\
\hline
1.30               &2.64 $\pm$0.25&--&(1.6$\pm$0.4)\E{2}\\
1.36               &4.9$\pm$1.1   &660$\pm$100 &(2.0$\pm$0.3)\E{2}\\
1.45               &16$\pm$2    &--&(3.5$\pm$0.8)\E{2}\\
\hline
\end{tabular}
\end{table}

\begin{table}
\caption[tab1]{\label{tab:1}  The results for  the ratio of the cross  sections
  of the
\pptpi\  reaction and the \ppe\ reaction with the corresponding \et\ decay 
into $3\pi$. Both statistical and systematical uncertainties are shown. 
In addition the extracted cross sections for \pppopopo\ and \pppipipi\ 
reactions are listed.
 }
\begin{tabular}{rrr}
\hline
Beam kinetic&$pp\pi^0\pi^0\pi^0$&  $\sigma(\pppopopo)$\\
energy $[$MeV$]$ &ratio prompt/\et&   $[\mu$b$]$        \\
\hline
1299$\pm$2& 0.480$\pm$0.053$\pm$0.11         &$0.42\pm 0.05\pm0.10$\\
1361$\pm$2& 0.550$\pm$0.014$^{+0.06}_{-0.09}$&$0.89\pm  0.02\pm 0.23$\\
1448$\pm$3& 0.260$\pm$0.010$^{+0.15}_{-0.08}$&$1.34\pm  0.05^{+0.80}_{-0.45}$\\
\hline
\hline
Beam kinetic&$pp\pi^+\pi^-\pi^0$&$\sigma(\pppipipi)$\\
energy $[$MeV$]$&ratio prompt/\et   & $[\mu$b$]$\\
\hline
1361$\pm$2& 4.1$\pm$0.3$\pm$0.4&$4.6\pm 0.3\pm 1.2$\\
\hline
\end{tabular}
\end{table}

\section{Results and discussion}

The main result of the experiment is the ratio between the \pptpi\ and
$pp \rightarrow pp\eta$  cross sections for which values  are given in
table~\ref{tab:1}  with  statistical  and systematical  uncertainties.
One  can  extract the  total  cross  section  values using  the  known
$\sigma_{pp\eta}$  from table~\ref{tab:0}  together with  the relevant
\et\ decay branching ratio into  $3\pi$. For the beam proton energy at
1.30  GeV the  \et\ cross  section points  come from  four experiments
\cite{Hibou:1998de,  Moskal:2003gt,  Calen:1996mn,  Chiavassa:1994ru}.
At  1.36 GeV  there  is  only one  data  point \cite{Calen:1996mn}  to
compare  with.    The  cross  section   value  at  1.45  GeV,   is  an
interpolation using the  data from PINOT \cite{Chiavassa:1994ru}.  The
extracted values  of cross  sections for the  production of  the three
pions are also given in table~\ref{tab:1}.

As a cross check of the acceptance corrections for the \ppe\ reaction,
the  cross sections  have  also been  estimated  using the  luminosity
derived  from  the  simultaneously  measured $pp$  elastic  scattering
events  ~\cite{Demirors:2005aa}.  The  scattering cross  sections were
taken from  the precision experiment  EDDA ~\cite{Albers:2004iw}.  The
obtained values  for the three energies  1.30, 1.36 and  1.45 GeV are:
3.36$\pm$0.17$\pm$0.5   $\mu$b,   5.06$\pm$0.26$\pm$0.6   $\mu$b   and
14.9$\pm$0.75$\pm$1.0 $\mu$b  respectively.  The uncertainties include
statistics,   luminosity  determination   (5\%)  and   the  acceptance
correction (including  uncertainty of the reaction model).   We find a
general agreement with  the $\sigma_{pp\eta}$ results from literature,
which indicates that detection acceptances are well under control.

\begin{figure}
\begin{center}
\includegraphics[width=0.85\textwidth,clip]{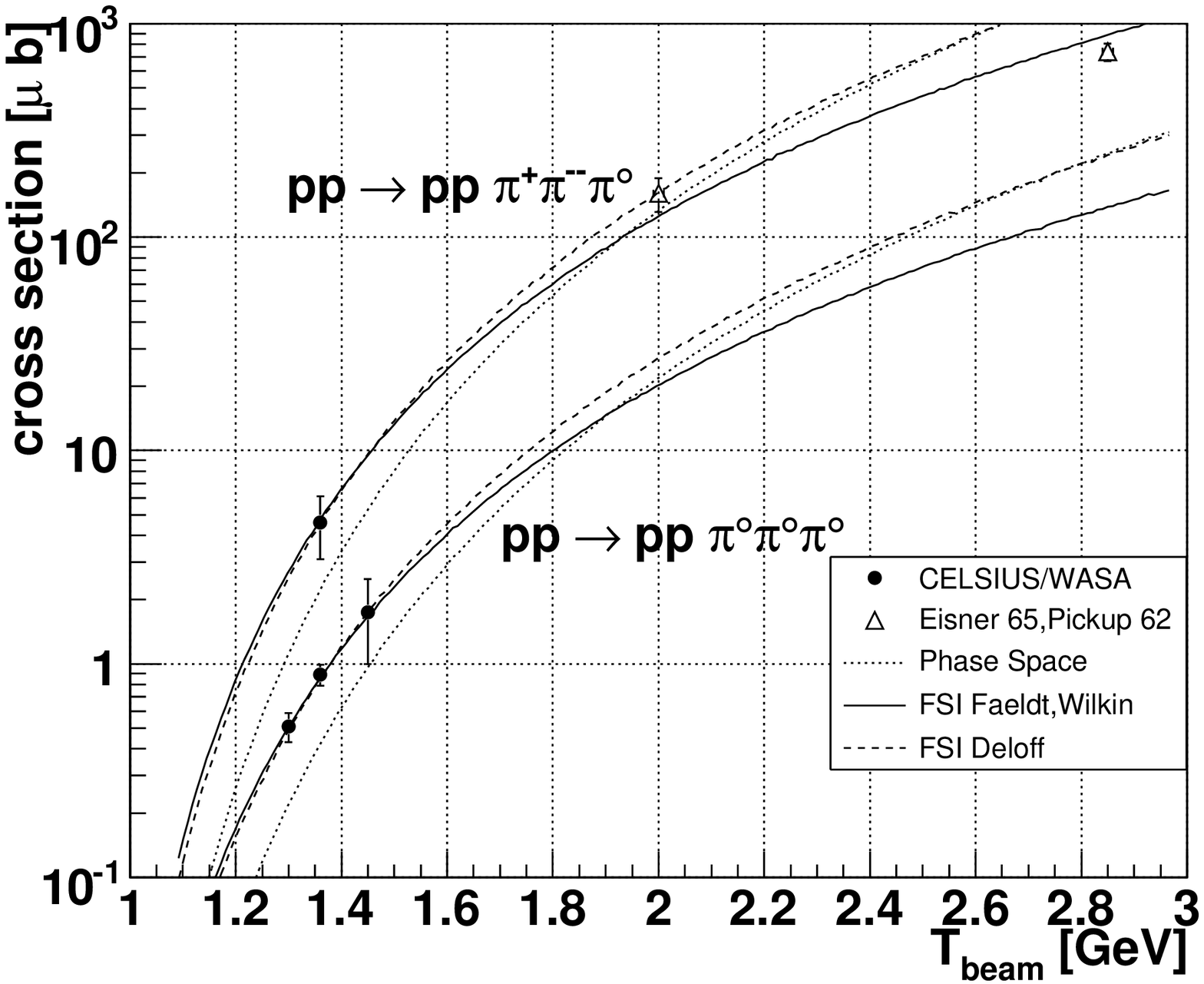}
\caption[pi1]{\label{fig:csection1} Measurements of the \pppipipi\ and
\pppopopo\ cross  section in  the beam proton  energy range  1--3 GeV.
The  four experimental  points  below  1.5 GeV  are  from the  present
experiment, while the data at 2 GeV and at 2.85 GeV for the \pppipipi\
reaction      are       from      bubble      chamber      experiments
\cite{Pickup:1962aa,Hart:1962aa}.  Dotted lines are predictions of the
statistical model  \cite{Fermi:1950jd} (phase space  and flux factor).
Dashed  and solid  line  predictions  are with  $pp$  FSI from  Deloff
\cite{Deloff:2003te}  and from F\"aldt  and Wilkin\cite{Faeldt:1996na}
respectively.}
\end{center}
\end{figure}

Figure~\ref{fig:csection1}  summarizes the  results for  the reactions
\pppipipi\  and  \pppopopo.  The  solid  lines  include  the $pp$  FSI
calculated according  to the parameterization from  F\"aldt and Wilkin
\cite{Faeldt:1996na} and the dashed lines the $pp$ FSI calculated with
the Reid  wave function \cite{Deloff:2003te} (here,  for $pp$ relative
momenta in center of mass greater  than 300 MeV/c, pure phase space is
used).  The  lines from  the model predictions  are normalized  to the
experimental data  points at 1.36  GeV.  In addition the  dotted lines
give the  energy dependence of  the cross section calculated  from the
statistical model  and are normalized  at high energies to  the dashed
lines.  Two data points for \pppipipi\ from bubble chamber experiments
in the energy range 2--3 GeV \cite{Pickup:1962aa,Hart:1962aa} are also
shown.   The  data  point  at  2  GeV  \cite{Pickup:1962aa}  has  been
corrected  by subtracting  the fraction  of events  expected  from the
$pp\rightarrow pp\et$ reaction \cite{Pickup:1962ab}.

The experimental  value obtained for  the ratio of the  cross sections
for     \pppipipi\      and     \pppopopo\     at      1.36~GeV     is
5.2$\pm$0.5(stat)$\pm$0.8(syst).   To have  a comparison  at  the same
$Q$, the  result should  be corrected for  the difference  between the
charged and neutral pion masses, which makes the phase space volume at
1.36 GeV for  the \pppopopo\ reaction 18\% larger  than for \pppipipi.
The      threshold     corrected      value     for      the     ratio
$\sigma(\pppipipi)/\sigma(\pppopopo)$           thus           becomes
6.3$\pm$0.6$\pm$1.0.  As mentioned in the introduction a similar ratio
is  expected in  the statistical  approach with  all  amplitudes being
equal.   The simplest  two  baryon excitation  mechanism involves  the
formation of  an intermediate $\Delta$  and an $N^*(1440)$,  where the
$N^*(1440)$ decays either to  $N(\pi\pi)^{T=0}_{s-\rm wave}$ or via an
intermediate    $\Delta\pi$   state    \cite{Yao:2006px}.     In   the
$N^*(1440)\to N(\pi\pi)^{T=0}_{s-  \rm wave}$ scenario  $T_{3\pi}$ has
to be 1.  Then only the amplitude $M_{111}$ can contribute and a cross
section ratio of  4 is expected from the  expressions given in section
1.  This decay branch can thus certainly  be a major part of the total
reaction mechanism.   For the scenario  $N^*(1440)\to\Delta\pi$, $T$ =
0,1 or 2 is permitted, and thus all $M_{101}$, $M_{111}$ and $M_{121}$
amplitudes can be involved and the ratio for the $3\pi$ system can not
be calculated without further  assumptions.  Note however that in this
case  the experimental  ratio  can easily  be  reproduced by  choosing
certain values of the amplitudes $M_{101}$ and $M_{121}$.  For example
the  choice $|M_{101}|^2\approx  0.7 |M_{111}|^2$  and $|M_{121}|^2=0$
gives the observed ratio  6.3.  Accordingly, also the reaction diagram
involving $N^*(1440)\to\Delta\pi$  might well  be the leading  part of
the  reaction  mechanism.  This  conclusion  is  further supported  by
calculations  within the  isobar model  by Sternheimer  and Lindenbaum
\cite{Sternheimer:1961}  which  lead  to  the  ratio 7.   In  case  of
$N(1520)$  in the intermediate  state one expects a much larger value,
since the  additional $N\rho$  decay mode can  contribute only  to the
\pppipipi\ channel.

Comparison of  the presented result  with the cross  sections obtained
from the Crystal Ball  data \cite{Starostin:2003cc} on $\pim p\to 3\po
n$ allows  to rule out  a mechanism with sequential  $N(1535)$ decays.
In  both  $pp$  and  $\pim  p$ interactions  the  \et\  production  is
dominated by  the excitation of  the $N(1535)$ resonance and  close to
the thresholds one expects  that $\sigma(\pim p\to 3\po n)/\sigma(\pim
p\to \eta  n)\approx \sigma(pp\to pp3\po)/\sigma(p p\to  pp\eta )$ for
this   scenario.    However,   the   ratio  $\sigma(\pim   p\to   3\po
n)/\sigma(\pim  p\to \eta  n)$ was  measured to  be below  one percent
\cite{Starostin:2003cc},  much lower than  the 10\%  for $\sigma(pp\to
pp3\po)/\sigma(p p\to pp\eta )$ presented in this paper.

In conclusion microscopic model calculations, of the same kind as
those existing for the double pion production
\cite{Alvarez-Ruso:1997mx}, are needed to shed more light on the issue
and exploit the result of the presented measurements. Experimentally,
more data are desired to get information on the cross sections for the
other three pion reaction channels.

\section{Acknowledgments}

We are grateful to the  personnel at The Svedberg Laboratory for their
support during the  course of the experiment.  We  would like to thank
Andrzej Deloff  for giving  access to his  computer code for  $pp$ FSI
calculations.   This  work  has   been  supported  by  BMBF  (06HH152,
06TU261),  by  Russian  Foundation  for  Basic  Research  (Grant  RFBR
02-02-16957)  and  the   European  Community  Research  Infrastructure
Activity under FP6, Hadron Physics, RII-CT-2004-506078.

\end{document}